\documentclass[10pt,aps,pra,english,superscriptaddress,nofootinbib
]{revtex4-2}
\usepackage{slashed}
\usepackage[T1]{fontenc}
\usepackage[latin9]{inputenc}
\usepackage{color}
\usepackage{babel}
\usepackage{mathrsfs}
\usepackage{array}
\usepackage{multirow}
\usepackage{amssymb}
\usepackage{graphicx}
\usepackage{subcaption}
\usepackage{ragged2e}
\usepackage{tikz}
\usepackage{amsmath}

\makeatletter
\@ifundefined{showcaptionsetup}{}{%
 \PassOptionsToPackage{caption=false}{subfig}}
\makeatother

\usepackage[colorlinks=true, linkcolor=red, citecolor=blue, urlcolor=blue, hypertexnames=false]{hyperref}

\begin{document}

%\flushbottom

\title{ Lattice spectra of \textit{DDK} three-body system with Lorentz covariant kinematic}
\author{Qi-Chao Xiao}
\affiliation{College of Science, University of Shanghai for Science and Technology, Shanghai 200093, China}

\author{Jin-Yi Pang}\email{ jypang@usst.edu.cn}
\affiliation{College of Science, University of Shanghai for Science and Technology, Shanghai 200093, China}

\author{ Jia-Jun Wu}\email{wujiajun@ucas.ac.cn}
\affiliation{School of Physical Sciences, University of Chinese Academy of Sciences, Beijing 100049, China}
%\emailAdd{jypang@usst.edu.cn}

%\date{}

\begin{abstract}
The $DDK$ system has gain increasing attention in recent research due to its potential to contain a three-hadron bound state. This article utilizes an extension of the Non-Relativistic Effective Field Theory (NREFT) and the finite volume particle-dimer framework to derive Lorentz-invariant quantization conditions for the $DDK$ three-body system. Using current model input conditions, the finite volume energy spectrum of the $DDK$ three-body system was calculated. This new calculation incorporates relativistic kinematics, allowing it to be applicable across a broader energy range starting from the threshold. In this work, we present a comprehensive \( O(p^{2}) \) calculation. The spurious pole is effectively subtracted within the framework of relativistic kinematics. The spectra in the moving frame are also obtained. These analyses provide a broader testing ground for future lattice simulations. They are expected to reveal more detailed properties of the $DDK$ system and other three-hadron systems.
 \end{abstract}

%\allowdisplaybreaks
\maketitle

\section{Introduction}\label{sec1}
Whether $DDK$ three-body system can form a bound state is an interesting
problem that has received a lot of attentions [\citenum{Sanchez:2018cp}-\citenum{Huang:2020cp}]. The Ref.\cite{Wu:2019cp} found that a $DDK$ bound state exists with
a binding energy of about $70\text{ MeV}$ due to the attractive $DK$
interaction. In this scenario, the quantum numbers of the three-body
bound state are $J^{P}=0^{-}$, $I=1/2$, $S=1$ and $C=2$, which
is an explicitly exotic state. Lattice QCD is an important tool to
solve the problem and verify this statement.  However, to achieve it,
not only are three-body lattice simulations required, but also the
corresponding finite volume analysis in order to extract the relevant
three-body observables from the lattice spectra. Recently, significant
progress has been made in both two sides [\citenum{Hansen2014}-\citenum{Doring:2018cp}].  Among these works, there are mainly three different but conceptually
equivalent formalisms known as Relativistic Field Theory (RFT) \cite{Hansen2014,Hansen2015}, Non-Relativistic Effective Field Theory (NREFT) \cite{Hammer:2017cp,Hammer2:2017cp} and Finite-Volume Unitarity (FVU) \cite{Mai2017,Mai2019} approches. In this paper, we are using NREFT to generate $DDK$'s
lattice spectra based on the above phenomenological calculations.
The idea was implemented in our previous paper \cite{ddk}, but
we used non-relativistic kinematics at that time. As the momentum
of light mesons, e.g., $K$ increases, especially in the moving frame,
non-relativistic kinematics will no longer be applicable. Therefore,
based on our recently developed Lorentz-invariant NREFT method \cite{NREFT},
we can extend our previous work \cite{ddk} to the relativistic
case. By doing so, we can not only extend to higher energy regions
but also analyze the spectra in the moving frames, making LQCD studies
of the $DDK$ system more efficient. Additionally, our approach may
be applied to the $DD\pi$ or $D\bar{D}\pi$ system, where the kinematics
of $\pi$ is necessarily relativistic. It has been known that lattice
studies of $DD\pi$ are of significant importance for $T_{cc}(3875)$
\cite{Hansen2024}, so is $D\bar{D}\pi$ for $\chi_{c1}(3872)$.

In order to generate $DDK$'s lattice spectra in a Lorentz covariant
manner, we need the corresponding particle-dimer formalism with relativistic
kinematics in the infinite volume firstly. This is a generalization
of \cite{NREFT} by considering two channels, i.e., $DK$
dimer with spectator $D$ and $DD$ dimer with spectator $K$. In
the particle-dimer scatteing equation, both dimer propagators should
be analyticly continuated below threshold. It has been known that
there is spurious pole \cite{fakepole} in dimer propagator at
$O(p^{2})$. How to subtract it from $DK$ or $DD$ dimer in a Lorentz-invariant
manner while preserving analyticity and unitarity is an issue to be
addressed in this paper. At the same time, we also need to introduce
higher order three-body couplings, e.g., $H_{2}$, in addition to
the leading $H_{0}$, to parameterized the interaction of $DDK$ in
coordination with effective range expansion (ERE) at $O(p^{2})$ in
two-body sector. By doing so, we can generate lattice spectra of $DDK$
in various moving frames through Lorentz covariant quantization condition
equipped with these three-body couplings. The spectra are expected
to compare with the future lattice simulations, to check whether $DDK$
bound state exists from the first principle. 

The paper is organized as follows. In Sec.\ref{sec:Effective-field-theory},
we establish the Lorentz-invariant effective field theory of $DDK$
system in the infinite volume based on \cite{NREFT}. We apply
the Lorentz-invariant particle-dimer formalism in a finite volume
in Sec.\ref{sec:DDK-finite-vol}. Quantization condition is given
from the covariant particle-dimer scattering equation and projected
onto irreps of $O_{h}$ for the rest frame and little
groups for the corresponding moving frames. In Sec.\ref{sec:Results-and-discussions},
we present the numerical calculations in both infinite and finite
volume. In the infinite volume, the three-body coupling constants
of $DDK$ system are fitted and logarithmic periodical behavior is
observed. In a finite volume, we show the lattice spectra of $DDK$
in the rest frame and moving frames within a Lorentz covariant formalism.
At last, the conclusion is given in Sec.\ref{sec4}.
\section{Effective field theory of \textit{DDK} system}\label{sec:Effective-field-theory}
\subsection{Lorentz-invariant particle-dimer Lagrangian}
To establish the quantization condition for the $DDK$ system in a
finite volume, we first need to develop an appropriate effective field
theory in the infinite volume. Here, we formulate the effective Lagrangian
for the $DDK$ system within the framework of Lorentz-invariant non-relativistic
effective field theory (NREFT) as proposed in \cite{NREFT}. Utilizing
the particle-dimer approach, we consider the three-body system to
be described by a $DK$ dimer with a spectator $D$ meson, as well
as a $DD$ dimer with a spectator $K$ meson. In this context, the
Lagrangian consists of three components: the kinematic terms $\mathcal{L}_{1}$
for the single particles ($D$ and $K$ mesons) and the dimers ($DK$
and $DD$); the coupling terms between a pair of two particles and
the dimer that they form, denoted as $\mathcal{L}_{2}$; and the interaction
terms between the dimer and the spectator particle, denoted as $\mathcal{L}_{3}$,
\begin{equation}
\mathcal{L}=\mathcal{L}_{1}+\mathcal{L}_{2}+\mathcal{L}_{3}.
\end{equation}
Firstly, the kinematic part is given by 
\begin{align}
\mathcal{L}_{1} & =D^{\dagger}2w_{v}\big(v\cdot i\partial-w_{v}\big)D+K^{\dagger}2w_{v}\big(v\cdot i\partial-w_{v}\big)K\nonumber \\
 & +\sigma_{DK}T_{DK}^{\dagger}T_{DK}+\sigma_{DD}T_{DD}^{\dagger}T_{DD}.
\end{align}
We denote the $DK$ dimer field by $T_{DK}$ and the $DD$ dimer field
by $T_{DD}$. These are auxiliary fields that encode the two-body
dynamics. In this work, we adopt relativistic kinematics, where $v^{\mu}$
is a unit timelike vector. When $v^{\mu}=(1,0,0,0)$, it indicates
that we are in the rest frame. Due to Lorentz invariance, all choices
of $v^{\mu}$ are physically equivalent. We then define the differential
operator $w_{v}$ as follows, 
\begin{equation}
w_{v}=\sqrt{m^{2}+\partial^{2}-(v\cdot\partial)^{2}}.
\end{equation}
It is a covariant representation of on-shell energy, saying, $w_{v}\to\omega=\sqrt{m^{2}+|\mathbf{p}|^{2}}$
for $v^{\mu}=(1,0,0,0)$. $\sigma$ is the bare parameter of the dimer
field. Secondly, the two-body part is 
\begin{align*}
\mathcal{L}_{2} & =T_{DK}^{\dagger}(D\mathscr{F}_{DK}K)+\frac{1}{2}T_{DD}^{\dagger}(D\mathscr{F}_{DD}D)+\text{h.c.}
\end{align*}
Here $\mathscr{F}$ is differential operator which represents the
interaction between the two particles inside the dimer. Up to $O(p^{2})$,
we have 
\begin{align*}
D\mathscr{F}_{DD}D & =DD+\frac{1}{8}f_{DD}\big(D\bar{w}_{\perp}^{\mu}\bar{w}_{\perp\mu}D-(\bar{w}_{\perp}^{\mu}D)(\bar{w}_{\perp\mu}D)\big).
\end{align*}
To understand this differential operator, we need to introduce firstly,
\begin{align}
w^{\mu} & =v^{\mu}w_{v}+i\partial_{\perp}^{\mu},\text{ where }\partial_{\perp}^{\mu}=\partial^{\mu}-v^{\mu}(v\cdot\partial).
\end{align}
Considering the boost $\Lambda_{\nu}^{\mu}$ that renders the total
four-momentum of the pair parallel to the vector $v^{\mu},$ we build
up that 
\begin{align}
\bar{w}^{\mu} & =\Lambda_{\nu}^{\mu}w^{\nu}.
\end{align}
Its space-like component gives the $\bar{w}_{\perp}^{\mu}$ at last,
that is 
\begin{align}
\bar{w}_{\perp}^{\mu} & =\bar{w}^{\mu}-v^{\mu}(v\cdot\bar{w}).\label{eq999}
\end{align}
 For inequal-mass case, i.e., $DK$-dimer, we have, additionally,

\begin{align*}
D\mathscr{F}_{DK}K & =DK+\frac{1}{8}f_{DK}\big(u_{D}^{2}D\bar{w}_{\perp}^{\mu}\bar{w}_{\perp\mu}K+u_{K}^{2}K\bar{w}_{\perp}^{\mu}\bar{w}_{\perp\mu}D\\
 & -2u_{D}u_{K}(\bar{w}_{\perp}^{\mu}D)(\bar{w}_{\perp\mu}K)\big).
\end{align*}
 Here,$u_{D(K)}$ generates $1\pm2(m_{D}^{2}-m_{K}^{2})/s_{DK}$ for
DK-dimer in the momentum space. Finally, the three-body Lagrangian
is given by

\begin{align}
\mathcal{L}_{3} & =T_{DK}^{\dagger}D^{\dagger}\mathscr{H}T_{DK}D,
\end{align}\label{eq10}
where $\mathscr{H}$ is operator that gives the three-body coupling
in the form of, 
\begin{align}
\mathscr{H} & =h_{0}+h_{2}\Delta.
\end{align}
In the momentum space, the operator $\Delta$ generates $s-s_{\text{th}}$,
where $s_{(\text{th})}$ is invariant mass squared of the three-body
system (at threshold). $h_{0}$ and $h_{2}$ are three-body coupling
constants between dimer and the spectating particle. Here, we only
parameterize the three-body force in one channel $(DK)+D\to(DK)+D$.
It seems to be more complete that we need to do it for all possible
channels, e.g., $(DK)+D\to(DD)+K$ and so on, that is
\begin{align}
T_{DK}^{\dagger}D^{\dagger}\mathscr{H}T_{DK}D & +T_{DD}^{\dagger}K^{\dagger}\mathscr{H}^{'}T_{DD}K.\nonumber \\
 & +T_{DD}^{\dagger}K^{\dagger}\mathscr{H}^{''}T_{DK}D+\text{h.c.}
\end{align}
As a matter of fact, considering the equivalent effective field theory
without dimers, we have only one operator for $DDK$ three-body interaction,
that is $(D^{\dagger}D^{\dagger}K^{\dagger})\mathscr{H}_{0}(DDK)$.
It means, we have the relationship between $\mathscr{H}_{0}$ and
$\mathscr{H}$, $\mathscr{H}^{'}$ and $\mathscr{H}^{''}$ in particle-dimer
formalism, 
\begin{align}
\mathscr{H}_{0} & =\frac{\mathscr{F}_{DK}\mathscr{H}\mathscr{F}_{DK}}{\sigma_{DK}^{2}}+\frac{\mathscr{F}_{DD}\mathscr{H}^{'}\mathscr{F}_{DD}}{\sigma_{DD}^{2}}+\frac{2\mathscr{F}_{DD}\mathscr{H}^{''}\mathscr{F}_{DK}}{\sigma_{DD}\sigma_{DK}}.
\end{align}
Apparently, we are allowed to choose $\mathscr{H}^{'}=\mathscr{H}^{''}=0$
and use the contact term between $DK$ dimer and the spectating $D$
to parameterize the three-body interaction, as we have discussed in \cite{ddk}

\subsection{Dimer propagator}

We are going to derive the dimer propagator in this section. Let us begin with the single particle's propagator, that is 
\begin{align}
 i \int dx \, e^{ikx} \langle T \phi^+(x) \phi(0) \rangle = \frac{1}{2w_{v}(k)(w_{v}(k)-v\cdot k-i\epsilon)}. 
\end{align}
Here, $\phi$ can represent either $D$ or $K$. In order to construct
the full propagator of the dimer, we need to calculate the self-energies
of $DK$ and $DD$ dimer, which are represented by the following loop
integrals, 
\begin{align}
I_{DK} & =\int\frac{d^{4}k}{(2\pi)^{4}i}\frac{F_{DK}^{2}(\bar{k}_{\perp}^{2})}{2w_{v}^{K}(k)(w_{v}^{K}(k)-v\cdot k-i\epsilon)}\nonumber \\
 & \times\frac{1}{2w_{v}^{D}(P-k)(w_{v}^{D}(P-k)-v\cdot(P-k)-i\epsilon)}.
\end{align}
We use different superscripts $D$ and $K$ in $w_{v}$, which means,
$w_{v}^{D(K)}(k)=\sqrt{m_{D(K)}^{2}-k^{2}+(v\cdot k)^{2}}$. $P$
is the dimer's total momentum and $s=P^{2}$. And $\bar{k}_{\perp}$
is defined in the same manner as $\bar{w}_{\perp}$in eq.(\ref{eq999}).
Since the vertex function is truncated up to $O(p^{2})$, we have,
\begin{align}
F_{DK}(k) & =1+\frac{1}{2}f_{DK}k^{2}.
\end{align}
By using the threshold expansion \cite{NREFT}, we can find that 
\begin{align}
I_{DK} & =\int\frac{d^{4}k}{(2\pi)^{4}i}\frac{F_{DK}^{2}((p_{DK}^{*})^{2})}{(m_{K}^{2}-k^{2}-i\epsilon)}\frac{1}{(m_{D}^{2}-(P-k)^{2}-i\epsilon)}\nonumber \\
 & =\big(1+f_{DK}(p_{DK}^{*})^{2}\big)\Sigma_{DK}(s).
\end{align}
Here $p_{DK}^{*}$ is on-shell relative momentum in $DK$ dimer, that
is $p_{DK}^{*}=\lambda^{1/2}(s,m_{D}^{2},m_{K}^{2})/(2\sqrt{s})$, and we always have $\text{Im}p_{DK}^{*}\geq0$.
$\Sigma_{DK}$ is Chew-Mandelstam functions, 
\begin{align}
\Sigma_{DK}(s) & =\frac{1}{16\pi^{2}}\Big(\frac{2p_{DK}^{*}}{\sqrt{s}}\log\frac{m_{D}^{2}+m_{K}^{2}-s+2p_{DK}^{*}\sqrt{s}}{2m_{D}m_{K}}\nonumber \\
 & -(m_{D}^{2}-m_{K}^{2})\big(\frac{1}{s}-\frac{1}{(m_{D}+m_{K})^{2}}\big)\log\frac{m_{D}}{m_{K}}\Big).\label{eq:CM-func}
\end{align}
By the same token, the self-energy of $DD$ dimer is, 
\begin{align}
I_{DD} & =\frac{1}{2}\int\frac{d^{4}k}{(2\pi)^{4}i}\frac{F_{DD}^{2}(\bar{k}_{\perp}^{2})}{2w_{v}^{D}(k)(w_{v}^{D}(k)-v\cdot k-i\epsilon)}\nonumber \\
 & \times\frac{1}{2w_{v}^{D}(P-k)(w_{v}^{D}(P-k)-v\cdot(P-k)-i\epsilon)}\nonumber \\
 & =\frac{1}{2}\int\frac{d^{4}k}{(2\pi)^{4}i}\frac{F_{DD}^{2}((p_{DD}^{*})^{2})}{(m_{D}^{2}-k^{2}-i\epsilon)}\frac{1}{(m_{D}^{2}-(P-k)^{2}-i\epsilon)}\nonumber \\
 & =\frac{1+f_{DD}(p_{DD}^{*})^{2}}{2}\Sigma_{DD}(s).
\end{align}
 Notice that $1/2$ is the symmetry factor of $DD$ system. The Chew-Mandelstam
function can be easily obtained by replacing $m_{K}$ in eq.(\ref{eq:CM-func})
with $m_{D}$. Based on the loop integral, the dimer propagator can
be calculated iteratively as follows,
\begin{align}
\tau_{DK}(s) & =\frac{-1}{\sigma_{DK}}+\frac{-1}{\sigma_{DK}}I_{DK}\frac{-1}{\sigma_{DK}}+\cdots\nonumber \\
 & =\frac{1}{-\sigma_{DK}-I_{DK}}\nonumber \\
 & =\left(\frac{1}{1+f_{DK}(p_{DK}^{*})^{2}}\right)\frac{1}{-\sigma_{DK}\big(1-f_{DK}(p_{DK}^{*})^{2}\big)-\Sigma_{DK}}.\label{eq:tau-pre}
\end{align}
The prefactor in the bracket can be dropped by redefining the particle-dimer amplitude and three-body parameters in Lagrangian, saying $h\rightarrow H$
\cite{Hammer2:2017cp}, so we are allowed to say 
\begin{align}
\tau_{DK}(s) & =\frac{1}{-\sigma_{DK}\big(1-f_{DK}(p_{DK}^{*})^{2}\big)-\Sigma_{DK}}.
\end{align}
Above the threshold, i.e., \( s \geq (m_{D} + m_{K})^{2} \), the Chew-Mandelstam function is 
\begin{align}
\Sigma_{DK} & =\text{Re}\Sigma_{DK}+\frac{ip_{DK}^{*}}{8\pi\sqrt{s}}.
\end{align}
So the dimer propagator can be expressed in the form of, 
\begin{align}
\text{\ensuremath{\tau_{DK}}} & =\frac{8\pi\sqrt{s}}{8\pi\sqrt{s}(-\sigma_{DK}\big(1-f_{DK}(p_{DK}^{*})^{2}\big)-\text{Re}\Sigma_{DK})-ip_{DK}^{*}}.
\end{align}
By the same token, we can find the $DD$ dimer's propagator, 
\begin{align}
\text{\ensuremath{\tau_{DD}}} & =\frac{16\pi\sqrt{s}}{16\pi\sqrt{s}(-\sigma_{DD}\big(1-f_{DD}(p_{DD}^{*})^{2}\big)-\text{Re}\Sigma_{DD}/2)-ip_{DD}^{*}}.
\end{align}
Here the additional factor $2$ is due to identical property. The
unitarity asks for that, up to $O(p^{2})$, 
\begin{align}
8\pi\sqrt{s}(-\sigma_{DK}\big(1-f_{DK}(p_{DK}^{*})^{2}\big)-\text{Re}\Sigma_{DK}) & \to p_{DK}^{*}\cot\delta_{DK},\label{eq:match-DK}\\
16\pi\sqrt{s}(-\sigma_{DD}\big(1-f_{DD}(p_{DD}^{*})^{2}\big)-\text{Re}\Sigma_{DD}/2) & \to p_{DD}^{*}\cot\delta_{DD}.
\end{align}
Using the effective range expansion (ERE), we can expand the following
term around the two-body threshold $s_{\text{th}}$, that is
\begin{align}
\frac{p\cot\delta}{8\pi\sqrt{s}}+\text{Re}\Sigma(s) & =b_{0}+b_{1}\delta s+O(\delta s^{2}),
\end{align}
where $\delta s=s-s_{\text{th}}$ and
\begin{align}
b_{0}=-\frac{1}{8\pi\sqrt{s_{\text{th}}}a}, & \;b_{1}=\frac{\pi a^{-1}+(\pi r\mu-2)\sqrt{s_{\text{th}}}+\zeta}{16\pi^{2}s_{\text{th}}^{3/2}}.\label{2828}
\end{align}
$\mu$ is the reduced mass of two particles inside the dimer. If the
two particles are identical, $\zeta=0$, otherwise, e.g., $DK$ dimer,
\begin{align}
\zeta & =(m_{D}-m_{K})\log\frac{m_{D}}{m_{K}}.
\end{align}
By using the above expansion, we can solve the match equation, that is, 
\begin{align}
-\sigma_{DK}\big(1-f_{DK}(p_{DK}^{*})^{2}\big) & =b_{0}^{(DK)}+b_{1}^{(DK)}\delta s,\label{eq:solved-match-DK}\\
-2\sigma_{DD}\big(1-f_{DD}(p_{DD}^{*})^{2}\big) & =b_{0}^{(DD)}+b_{1}^{(DD)}\delta s.
\end{align}
The solved match conditions finally give the dimer propagators in
the form of, 
\begin{align}
\tau_{DK}(s) & =\frac{1}{b_{0}^{(DK)}+b_{1}^{(DK)}(s-s_{\text{th}})-\Sigma_{DK}(s)},\\
\tau_{DD}(s) & =\frac{2}{b_{0}^{(DD)}+b_{1}^{(DD)}(s-s_{\text{th}})-\Sigma_{DD}(s)}.\label{eq:DD-unsub}
\end{align}
It turns out that these dimer propagators can be extended below threshold
naturally under the premise of maintaining analyticity and unitarity.

\subsection{Particle-dimer scattering equation}
\begin{figure}[h]
  \centering
  \includegraphics[width=\linewidth]{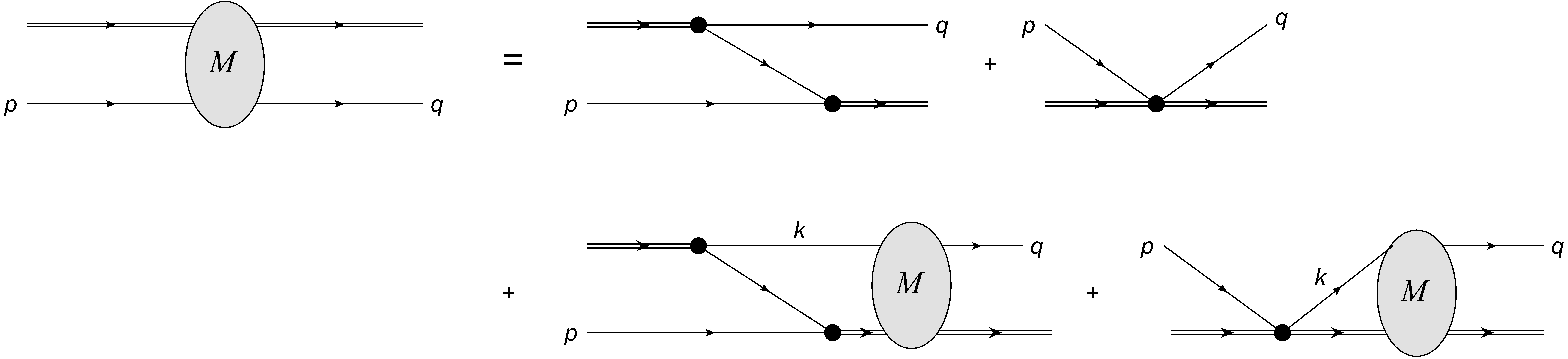}
  \caption{ Schematic 3-body scattering equation}
  \label{fig:1}
\end{figure}

For $DDK$ system, we can write down the scattering equation of particle-dimer
amplitude, $\mathcal{M}$ (see Fig.\ref{fig:1}) in the form of, 
\begin{align}
\mathcal{M}(p,q;P) & =Z(p,q;P)+\int\frac{d^{4}k}{(2\pi)^{4}}\Theta_{v}(k)Z(p,k;P)\tau((P-k)^{2})\mathcal{M}(k,q;P).\label{34}
\end{align}
Here, \( Z \) is the potential between the particle and the dimer. We use \( p \) and \( q \) to represent the momenta of the incoming and outgoing spectators, respectively, while \( k \) is for the spectator in the intermediate process. \( P \) denotes the total momentum of the \( DDK \) three-body system. The integral is truncated via the \( \Theta \) function, which takes a Lorentz covariant form,
\begin{align}
\Theta_{v}(k) & =2\pi\delta(k^{2}-m^{2})\theta(\Lambda^{2}+k^{2}-(v\cdot k)^{2}).\label{35}
\end{align}
Since there are two kinds of dimers, $\mathcal{M}$ and $Z$ are both $2\times2$ matrices, that is,
\begin{align}
\mathcal{M}=\begin{pmatrix}\mathcal{M}_{11} & \mathcal{M}_{12}\\
\mathcal{M}_{21} & \mathcal{M}_{22}
\end{pmatrix},\quad Z & =\begin{pmatrix}Z_{11} & Z_{12}\\
Z_{21} & Z_{22}
\end{pmatrix}.
\end{align}
For example, \( \mathcal{M}_{11} \) is the amplitude of \( (DK) + D \to (DK) + D \); correspondingly, \( Z_{11} \) represents the process of exchanging a \( K \) meson. Combining with the contact term in eq.(\ref{eq10}), we can write down the explicit form of \( Z_{11} \),
\begin{align}
Z_{11}(p,q;P) & =\frac{F_{DK}((\bar{p}_{\perp})^{2})F_{DK}((\bar{q}_{\perp})^{2})}{2w_{v}^{K}(P-p-q)\big(w_{v}^{K}(P-p-q)-v\cdot(P-p-q)\big)}\nonumber \\
 & +h_{0}+h_{2}(s-s_{\text{th}}),
\end{align}
with $s=P^{2}$ and $s_{\text{th}}$ is invariant mass squared at
the threshold. Using threshold expansion, we can put the off-shell
vertex information into the three-body force \cite{Hammer2:2017cp},
that is
\begin{align}
Z_{11}(p,q;P) & =(1+\frac{1}{2}f_{DK}(p_{DK}^{*})^{2})\Big\{\Big[2w_{v}^{K}(P-p-q)\times\nonumber \\
 & \times\big(w_{v}^{K}(P-p-q)-v\cdot(P-p-q)\big)\Big]^{-1}\nonumber \\
 & +\frac{H_{0}(\Lambda)}{\Lambda^{2}}+\frac{H_{2}(\Lambda)}{\Lambda^{4}}(s-s_{\text{th}})\Big\}(1+\frac{1}{2}f_{DK}(p_{DK}^{*})^{2}).
\end{align}
The two-body vertexes on the two sides cancel the same structure in
the neighbouring dimer propagator, so we express it as, 
\begin{align}
Z_{11}(p,q;P) & =\frac{1}{2w_{v}^{K}(P-p-q)\big(w_{v}^{K}(P-p-q)-v\cdot(P-p-q)\big)}\nonumber \\
 & +\frac{H_{0}(\Lambda)}{\Lambda^{2}}+\frac{H_{2}(\Lambda)}{\Lambda^{4}}(s-s_{\text{th}}).
\end{align}
For the process, $(DK)+D\to(DD)+K$, we use $\mathcal{M}_{12}$ to
denote the amplitude. This process is realized by the exchange of
$D$ meson, the potential $Z_{12}$ reads, 
\begin{align}
Z_{12}(p,q;P) & =\frac{1}{2w_{v}^{D}(P-p-q)(w_{v}^{D}(P-p-q)-v\cdot(P-p-q))}.
\end{align}
Apparently, due to time-reversal symmetry, $Z_{21}(p,q)=Z_{12}(q,p)$.
Since one cannot realize $(DD)+K\to(DD)+K$ via changing only one
particle, we have $Z_{22}=0$. In the scattering equation, dimer propagator
is diagonal, that is
\begin{align}
\tau & =\begin{pmatrix}\tau_{DK}\\
 & \tau_{DD}
\end{pmatrix}.
\end{align}
\subsection{Spurious pole}\label{sp}
In order to solve the particle-dimer scattering equation, it is necessary
to analyticly continuate the dimer propagator below threshold. It has
been known that ERE up to $O(p^{2})$ may generate spurious pole \cite{fakepole}. Although this unphysical effect does not affect two-body
physics, they can severely disrupt the unitarity of the three-body
equations. Therefore, in this section, we will present a method to
eliminate spurious poles within the Lorentz invariant framework. 
\par
Let us begin with the dimer propagator, 
\begin{align}
\tau(s) & =\frac{1}{b_{0}+b_{1}(s-s_{\text{th}})-\Sigma(s)}.
\end{align}
The spurious pole emerges when \( a > r > 0 \). In addition to the physical pole, we will also find a spurious pole \( s_{\text{f}} \) with an opposite residue \( R_{\text{f}} \). This spurious pole breaks unitarity and therefore must be subtracted. This can be achieved by \cite{fakepole},
\begin{align}
\tau & \to\tau-\frac{R_{\text{f}}}{s-s_{\text{f}}}+(\text{polynomial in }\delta s).
\end{align}
Here the polynomial is the expansion of $R_{\text{f}}/(s-s_{\text{f}})$, that is
\begin{align}
\frac{R_{\text{f}}}{s-s_{\text{f}}} & =c_{0}+c_{1}\delta s+O(\delta s^{2}).\label{eq:spurious expansion}
\end{align}
The coefficients are 
\begin{align}
c_{0} & =\frac{R_{\text{f}}}{s_{\text{th}}-s_{\text{f}}},\quad c_{1}=-\frac{R_{\text{f}}}{(s_{\text{th}}-s_{\text{f}})^{2}}.\label{4343}
\end{align}
Then the dimer propagator without the spurious pole reads ,
\begin{align}
\tau(s) & =\frac{1}{J(s)-\Sigma(s)},
\end{align}
where $J(s)$ is defined by 
\begin{align}
J(s) & =\Big[\frac{1}{b_{0}+b_{1}(s-s_{\text{th}})-\Sigma(s)}-\frac{R_{\text{f}}}{s-s_{\text{f}}}+c_{0}+c_{1}(s-s_{\text{th}})\Big]^{-1}+\Sigma(s).\label{eq88}
\end{align}
It is proved that this subtraction scheme can ensure unitarity and
analyticity at arbitrary order, as long as eq.(\ref{eq:spurious expansion}) is expanded up to sufficient high order. There is another possibility
that can generate spurious pole, i.e., $r>0>a$. For example, although
we have no $DD$ bound state, the unsubtracted dimer propagator, eq.(\ref{eq:DD-unsub})
also contains a spurious pole below threshold. Considering the symmetry
factor of $DD$ dimer, we can eliminate the spurious pole through
the following expression, 
\begin{align}
\text{\ensuremath{\tau}}_{DD} & =\frac{2}{J_{DD}-\Sigma_{DD}(s)},
\end{align}
with 
\begin{align}
J_{DD}(s) & =2\Big[\frac{2}{b_{0}^{(DD)}+b_{1}^{(DD)}(s-s_{\text{th}})-\Sigma_{DD}(s)}-\frac{R_{\text{f}}^{(DD)}}{s-s_{\text{f}}^{(DD)}}\nonumber \\
 & +c_{0}^{(DD)}+c_{1}^{(DD)}(s-s_{\text{th}})\Big]^{-1}+\Sigma_{DD}(s).
\end{align}
\section{\textit{DDK} system in a finite volume}\label{sec:DDK-finite-vol}
\subsection{Dimer propagator in a finite volume}
In a finite volume with cubic periodic boundary condition, particles'
momenta are discrete, saying,  $\mathbf{p}\in\left(2\pi/L\right)^{3}\mathbb{Z}^{3}$,
where $L$ is the spatial lattice size. For field theory, it means
all the loop integral should be replaced by loop summation. Let
us begin with the two-body loop of dimer propagator, 
\begin{align}
I_{DK}^{L}(P) & =\frac{1}{L^{3}}\sum_{\mathbf{k}}\int\frac{dk^{0}}{2\pi i}\frac{F_{DK}^{2}(\bar{k}_{\perp}^{2})}{2w_{v}^{D}(k)(w_{v}^{D}(k)-v\cdot k-i\epsilon)}\nonumber \\
 & \times\frac{1}{2w_{v}^{K}(P-k)(w_{v}^{K}(P-k)-v\cdot(P-k)-i\epsilon)}.
\end{align}
Here $P$ is momentum of dimer. We adopt sum-integral difference to
calculate $I^{L}$, that is 
\begin{align}
I_{DK}^{L} & =\text{Re}I_{DK}+\Delta_{DK}^{L}.
\end{align}
The finite volume correction $\Delta^{L}$ is defined by 
\begin{align}
\Delta_{DK}^{L}(P) & =\left(\frac{1}{L^{3}}\sum_{\mathbf{k}}-\text{PV}\int\frac{d^{3}k}{(2\pi)^{3}}\right)\int\frac{dk^{0}}{2\pi i}\frac{F_{DK}^{2}(\bar{k}_{\perp}^{2})}{2w_{v}^{D}(k)(w_{v}^{D}(k)-v\cdot k-i\epsilon)}\nonumber \\
 & \times\frac{1}{2w_{v}^{K}(P-k)(w_{v}^{K}(P-k)-v\cdot(P-k)-i\epsilon)}\nonumber \\
 & =\left(\frac{1}{L^{3}}\sum_{\mathbf{k}}-\text{PV}\int\frac{d^{3}k}{(2\pi)^{3}}\right)\int\frac{dk^{0}}{2\pi i}\frac{1+f_{DK}(p_{DK}^{*})^{2}}{(m_{D}^{2}-k^{2}-i\epsilon)(m_{K}^{2}-(P-k)^{2}-i\epsilon)}.
\end{align}
The last line is obtained by using threshold expansion within sum-integral
difference. After integrating out $k^{0}$, we can find that 
\begin{align}
\Delta_{DK}^{L} & =\big(1+f_{DK}(p_{DK}^{*})^{2}\big)S_{DK}^{L},
\end{align}
where
\begin{align}
S_{DK}^{L} & =\left(\frac{1}{L^{3}}\sum_{\mathbf{k}}-\text{PV}\int\frac{d^{3}k}{(2\pi)^{3}}\right)\frac{1}{4\omega_{D}(\mathbf{k})\omega_{K}(\mathbf{P}-\mathbf{k})}\frac{1}{\omega_{K}(\mathbf{P}-\mathbf{k})+\omega_{D}(\mathbf{k})-P^{0}},
\end{align}
where $\omega$ is on-shell energy, i.e., $\omega(\mathbf{k})=\sqrt{m^{2}+|\mathbf{k}|^{2}}$.
We can use L\"{u}scher zeta function to calculate $S^{L}$, that
is 
\begin{align}
S_{DK}^{L} & =\frac{1}{4\pi^{3/2}L\gamma\sqrt{s}}Z_{00}^{d}(1;\eta_{DK}^{2}).
\end{align}
Based on dimer's momentum $P$, we define invariant mass, $s=P^{2}$,
boost vector, $d=PL/(2\pi)$ and Lorentz factor, $\gamma=d^{0}/\sqrt{d^{2}}$.
For $DK$ dimer, the on-shell momentum is given by $\eta_{DK}=p_{DK}^{*}L/(2\pi)$.
The zeta function is calculated by
\begin{align}
Z_{00}^{d}(1;\eta^{2}) & =\frac{1}{\sqrt{4\pi}}\sum_{\mathbf{r}\in P_{d}}\frac{1}{r^{2}-\eta^{2}},
\end{align}
where $\mathbf{r}$ is valued in 
\begin{align}
P_{d} & =\big\{\mathbf{r}|\mathbf{r}_{\parallel}=\gamma^{-1}(\mathbf{n}_{\parallel}-\chi\frac{\mathbf{d}}{2}),\mathbf{r}_{\perp}=\mathbf{n}_{\perp},\mathbf{n}\in\mathbb{Z}^{3}\big\}.
\end{align}
Here, $\parallel$ and $\perp$ are both relative to $\mathbf{d}$.
The coefficient $\chi=1$ for identical two-body system, while in
non-identical two-body system, e.g., $DK$ system, it is 
\begin{align}
\chi & =1+\frac{m_{D}^{2}-m_{K}^{2}}{s}.
\end{align}
By the same token, the self energy of $DD$ dimer is 
\begin{align}
I_{DD}^{L} & =\text{Re}I_{DD}+\Delta_{DD}^{L},
\end{align}
with 
\begin{align}
\Delta_{DD}^{L} & =\frac{1}{2}\big(1+f_{DD}(p_{DD}^{*})^{2}\big)S_{DD}^{L},
\end{align}
and
\begin{align}
S_{DD}^{L} & =\frac{1}{4\pi^{3/2}L\gamma\sqrt{s}}Z_{00}^{d}(1;\eta_{DD}^{2}).
\end{align}
The additional factor $2$ is due to identical property. 

The dimer propagator in a finite volume is built up in the same way
as the infinite volume, where the only difference is in self energy.
So we have
\begin{align}
\tau_{DK}^{L} & =\frac{-1}{\sigma_{DK}}+\frac{-1}{\sigma_{DK}}I_{DK}^{L}\frac{-1}{\sigma_{DK}}+\cdots\nonumber \\
 & =\frac{1}{-\sigma_{DK}-I_{DK}^{L}}\nonumber \\
 & =\left(\frac{1}{1+f_{DK}(p_{DK}^{*})^{2}}\right)\frac{1}{-\sigma_{DK}\big(1-f_{DK}(p_{DK}^{*})^{2}\big)-\text{Re}\Sigma_{DK}-S_{DK}^{L}}.
\end{align}
The prefactor can be absorbed into three-body force, as we have seen
in eq.(\ref{eq:tau-pre}), so
\begin{align}
\tau_{DK}^{L} & =\frac{1}{-\sigma_{DK}\big(1-f_{DK}(p_{DK}^{*})^{2}\big)-\text{Re}\Sigma_{DK}-S_{DK}^{L}}.
\end{align}
Based on the match equation in the infinite volume, eq.(\ref{eq:solved-match-DK}),
$DK$ dimer's propagator is obtained as, 
\begin{align}
\tau_{DK}^{L}(P) & =\frac{1}{b_{0}^{(DK)}+b_{1}^{(DK)}(s-s_{\text{th}})-\text{Re}\Sigma_{DK}(s)-S_{DK}^{L}(P)}.
\end{align}
By the same token, $DD$ dimer's propagator reads, 
\begin{align}
\tau_{DD}^{L}(P) & =\frac{2}{b_{0}^{(DD)}+b_{1}^{(DD)}(s-s_{\text{th}})-\text{Re}\Sigma_{DD}(s)-S_{DD}^{L}(P)}.
\end{align}

The above finite volume propagators have spurious poles as well. We
can use the same trick in sec.(\ref{sp}) to eliminate
them, that is changing the dynamical term as
\begin{align}
b_{0}+b_{1}(s-s_{\text{th}}) & \to J(s).
\end{align}
So the finite volume propagators take the form at last, 
\begin{align}
\tau_{DK}^{L}(P) & =\frac{1}{J_{DK}(s)-\text{Re}\Sigma_{DK}(s)-S_{DK}^{L}(P)},\\
\tau_{DD}^{L}(P) & =\frac{2}{J_{DD}(s)-\text{Re}\Sigma_{DD}(s)-S_{DD}^{L}(P)}.
\end{align}
\subsection{Quantization condition}
The energy levels of $DDK$ system on the lattice is determined by
the poles of particle-dimer scattering amplitude $\mathcal{M}^{L}$
in a finite volume. This amplitude obeys the scattering equation,
\begin{align}
\mathcal{M}^{L}(p,q;P) & =Z(p,q;P)+\frac{1}{L^{3}}\sum_{\mathbf{k}}\tilde{\Theta}_{v}(k)Z(p,k;P)\tau^{L}((P-k)^{2})\mathcal{M}^{L}(k,q;P).
\end{align}
It is the finite volume version of eq.(\ref{34}),
where all the loop integrals are replaced by loop summations. Since
we have integrated out $k^{0}$, the truncation function $\tilde{\Theta}_{v}$
reads now, 
\begin{align}
\tilde{\Theta}_{v}(k) & =\frac{\theta(\Lambda^{2}+m^{2}-(v\cdot k)^{2})}{2\omega(\mathbf{k})}.
\end{align}
In a finite volume, we choose the velocity $v^{\mu}=P^{\mu}/\sqrt{P^{2}}$
to ensure the proper three-body coupling constants, $H_{0}$, $H_{2}$
are applied. The finite volume scattering equation generates quantization
condition of $DDK$ system, 
\begin{align}
\det\Big(\delta_{pq}-\frac{1}{L^{3}}\tilde{\Theta}_{v}(q)Z(p,q;P)\tau^{L}((P-q)^{2})\Big) & =0.
\end{align}
It can be projected with respect to the appropriate finite volume
symmetry \cite{Doring:2018cp}. For example, we have cubic symmetry $O_{h}$
for the rest frame and little groups for various moving frames. Firstly,
in the discrete momenta configuration, we define the shell whose element
is obtained by the symmetry transformation of the reference, that
is 
\begin{align}
r\text{-shell} & =\{\mathbf{p}|\mathbf{p}=g\mathbf{p}_{r}^{(0)},g\in G\}.
\end{align}
Here $G$ is the symmetry group and $\mathbf{p}_{r}^{(0)}$ is the
reference of $r$-shell. The potential $Z$ can be projected as
\begin{align}
Z(p,q) & =Z(gp_{r}^{(0)},g^{'}q_{r^{'}}^{(0)})=\sum_{\Gamma,\alpha\beta}\sum_{\lambda}\left(\mathscr{T}_{\beta\alpha}^{\Gamma}(g)\right)^{*}\frac{s_{\Gamma}}{|G|}Z_{\alpha\lambda}^{\Gamma}(r,r^{'})\Big(\mathscr{T}_{\beta\lambda}^{\Gamma}(g^{'})\Big).
\end{align}
$s_{\Gamma}$ is the dimension of the irreps $\Gamma$ and $\alpha,\beta,\lambda$
are indices of the basis in the representation space. $\mathscr{T}^{\Gamma}(g)$
is the representation matrix of group element $g$. Due to the symmetry,
the projected quantity is calculated as 
\begin{align}
Z_{\alpha\beta}^{\Gamma}(r,r^{'}) & =\sum_{g}\Big(\mathscr{T}_{\beta\alpha}^{\Gamma}(g)\Big)Z(gp_{r}^{(0)},q_{r^{'}}^{(0)}).
\end{align}
The finite volume amplitude $\mathcal{M}^{L}$ can be projected in
the same way. Then we have the projection for the scattering equation,
\begin{align}
\mathcal{M}_{\alpha\beta}^{\Gamma}(r,r^{'};P) & =Z_{\alpha\beta}^{\Gamma}(r,r^{'};P)+\frac{1}{L^{3}}\sum_{r^{''}}\frac{\vartheta_{r^{''}}}{|G|}\tilde{\Theta}_{v}(r^{''})\nonumber \\
 & \times\sum_{\gamma}Z_{\alpha\gamma}^{\Gamma}(r,r^{''};P)\tau((P-k_{r^{''}})^{2})\mathcal{M}_{\gamma\beta}^{\Gamma}(r^{''},r^{'};P).
\end{align}
We define $\vartheta$ as the shell multiplicity. Notice that $\tau$
needs not projection because it is invariant under the group transformation.
At last, the projected equation gives the projected quantization condition,
\begin{align}
\det\Big(\delta_{rr^{'}}\delta_{\alpha\beta}-\frac{1}{L^{3}}\frac{\vartheta_{r^{'}}}{|G|}\tilde{\Theta}_{v}(r^{'})Z_{\alpha\beta}^{\Gamma}(r,r^{'};P)\tau((P-k_{r^{'}})^{2})\Big) & =0.\label{eq:qc}
\end{align}
\section{RESULTS AND DISCUSSIONS}\label{sec:Results-and-discussions}
\subsection{Subtraction of spurious pole in dimer propagator}
\begin{figure}[h]
    \centering
    \begin{minipage}{0.47\textwidth}
        \centering
        \includegraphics[width=\linewidth]{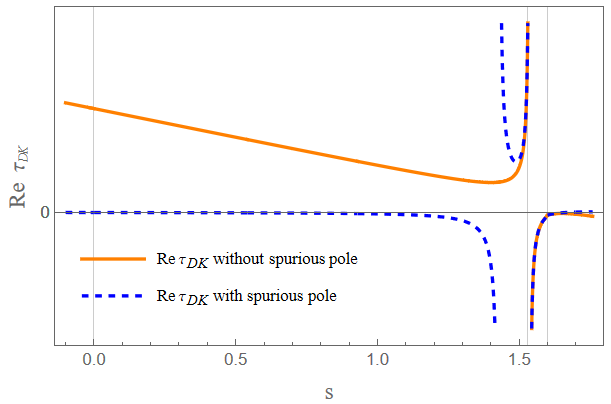}
    \end{minipage}\hfill
    \begin{minipage}{0.47\textwidth}
        \centering
        \includegraphics[width=\linewidth]{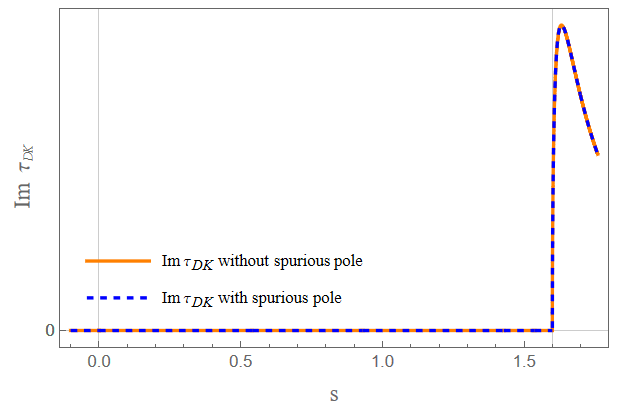}
        
    \end{minipage}
    
    \medskip
    
    \begin{minipage}{0.47\textwidth}
        \centering
        \includegraphics[width=\linewidth]{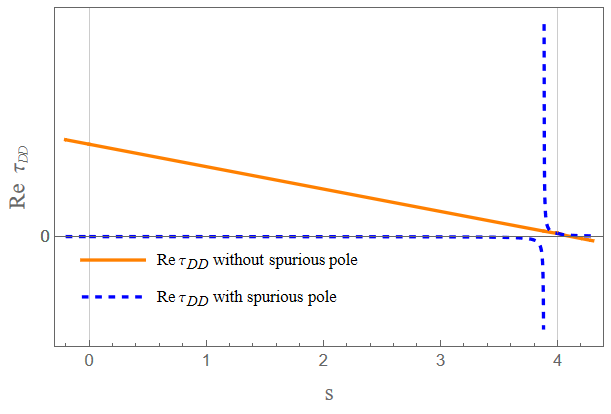}
       
    \end{minipage}\hfill
    \begin{minipage}{0.47\textwidth}
        \centering
        \includegraphics[width=\linewidth]{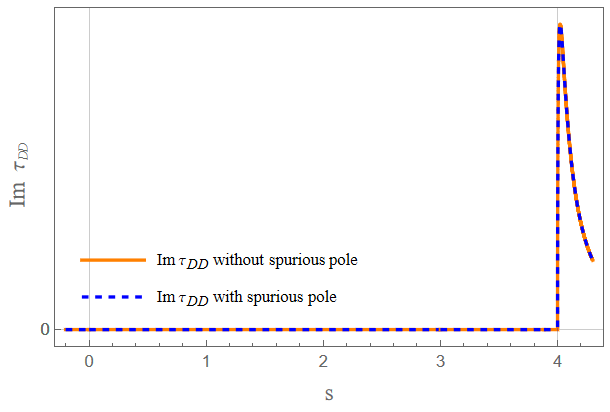}
       
    \end{minipage}
    
    \caption{\justifying Dimer propagators. All the quantities
are scaled by $m_{D}$. Upper left (right) is the real (imaginary)
part of $DK$ dimer; lower left (right) is the real (imaginary) part
of $DD$ dimer. In the real parts (left panel), spurious poles in blue dashed lines are eliminated, and dimer propagators without spurious poles are represented by the yellow line. The imaginary parts (right
panel) below threshold are always zero that means unitarity is preserved.
One can also see that the dimer propagator can be analyticly continuated
until $s<0$ in our formalism.}
    \label{fig:6}
\end{figure}

Since we are using ERE up to $O(p^{2})$, the dimer propagator is determined by the two-body scattering length and effective range. In our calculation, we use the same parameters as \cite{ddk}, see Table.\ref{tab1}. Applying eq.(\ref{2828}), we can write down the dimer propagator in the form of, 
\begin{align}
\tau(s) & =\frac{1}{b_{0}+b_{1}(s-s_{\text{th}})-\Sigma(s)}.\label{eq:spurious-prop}
\end{align}
\begin{table}[h]
\begin{centering}
\begin{tabular}{c|c|c}
\hline\hline
 & $a(\mathrm{fm})$ & $r(\mathrm{fm})$ \\
\hline
$DK$ & $1.683$ & $0.791$ \\
\hline
$DD$ & $-0.392$ & $3.236$ \\
\hline\hline
\end{tabular}
\par\end{centering}
\caption{Scattering lengths and effective ranges of DK and DD. These parameters are consistent with \cite{ddk}.}
\label{tab1}
\end{table}
For $DK$ dimer, there is a pole at $s=s^{*}$ which correspondes to $D_{s0}^{*}(2317)$, while for $DD$ dimer, no physical pole is observed. However, the spurious poles $s_{\text{f}}$ emerge in eq.(\ref{eq:spurious-prop}) for both $DK$ and $DD$ dimer. By changing problematic dynamical
terms $b_{0}+b_{1}(s-s_{\text{th}})$ to $J(s)$, we can eliminate the spurious poles while maintaining analyticity and unitarity, that is
\begin{align}
\tau(s) & =\frac{1}{b_{0}+b_{1}(s-s_{\text{th}})-\Sigma(s)}\to\frac{1}{J(s)-\Sigma(s)}.
\end{align}
This can be seen in Fig.\ref{fig:6} and the corresponding parameters are list
in Table.\ref{tab2}. 
\begin{table}[h]
\begin{centering}
\begin{tabular}{c|c|c|c|c|c|c|c}
\hline\hline
 & $b_{0}$ & $b_{1}$ & $s_{\mathrm{f}}$ & $s^{*}$ & $R_{\mathrm{f}}$ & $c_{1}$ & $c_{2}$ \\
\hline
$DK$ & $-0.0019$ & $0.0152$ & $1.4283$ & $1.5364$ & $276.0684$ & $1612.274$ & $-9415.88$ \\
\hline
$DD$ & $0.0053$ & $0.0723$ & $3.8834$ & $-$ & $33.1097$ & $284.077$ & $-2437.35$ \\
\hline\hline
\end{tabular}
\par\end{centering}
\caption{The parameters of the dimer propagators. All the parameters are scaled by $m_{D}$.}
\label{tab2}
\end{table}

\par
\begin{figure}[h]
  \centering
  \begin{subfigure}[b]{0.45\textwidth}
    \includegraphics[width=\textwidth]{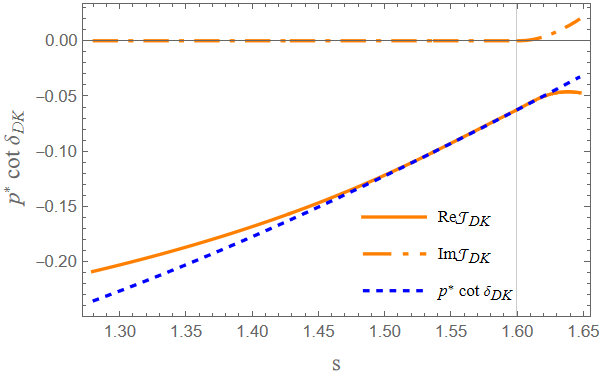}  
  \end{subfigure}
  \begin{subfigure}[b]{0.45\textwidth}
    \includegraphics[width=\textwidth]{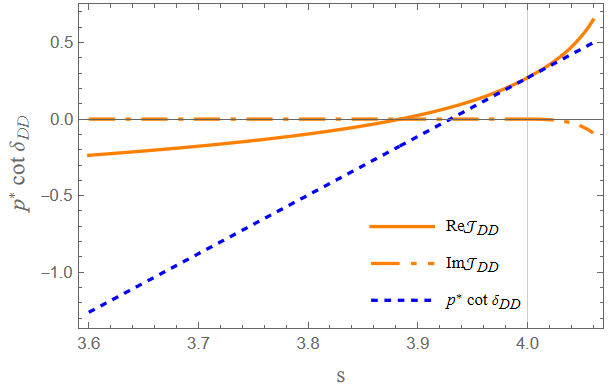}
  \end{subfigure}
  \caption{\justifying Analyticity and unitarity of dimer propagators. All the quantities
are scaled by $m_{D}$. Left panel is for $DK$ dimer, while right
panel is for $DD$. $\text{Im}\mathcal{J}$ represented by yellow dashed line always
vanishes at the given order, which maintains the unitarity. Above
the threshold, $\text{Re}\mathcal{J}$ representd by yellow line can reproduce
$p^{*}\cot\delta$ represented by  blue dashed line order by order. Below threshold,
$\text{Re}\mathcal{J}$ gives a more reasonable coninuation.}
  \label{fig:xy}
\end{figure}

To check the analyticity and unitarity in detail, we can rewrite the propagator in the form of, 
\begin{align}
\tau(s) & =\frac{8\pi\sqrt{s}}{8\pi\sqrt{s}[J(s)-\text{Re}\Sigma(s)]-ip^{*}}.
\end{align}
Let us define 
\begin{align}
\mathcal{J}(s) & =8\pi\sqrt{s}[J(s)-\text{Re}\Sigma(s)].
\end{align}
The unitarity is presented by $\text{Im}\mathcal{J}=0$. Above threshold, $\mathcal{J}(s)$ is consistent with phase shift term, i.e., $p^{*}\cot\delta$. Addtionally, it can provide a more resonable analytic form below threshold compared to the direct ERE. In our framework, this can be ensured order by order, as shown in Fig.\ref{fig:xy}. 

\subsection{Three-body coupling constants}
\begin{figure}[h]
  \centering
  \begin{subfigure}[b]{0.45\textwidth}
    \includegraphics[width=\textwidth]{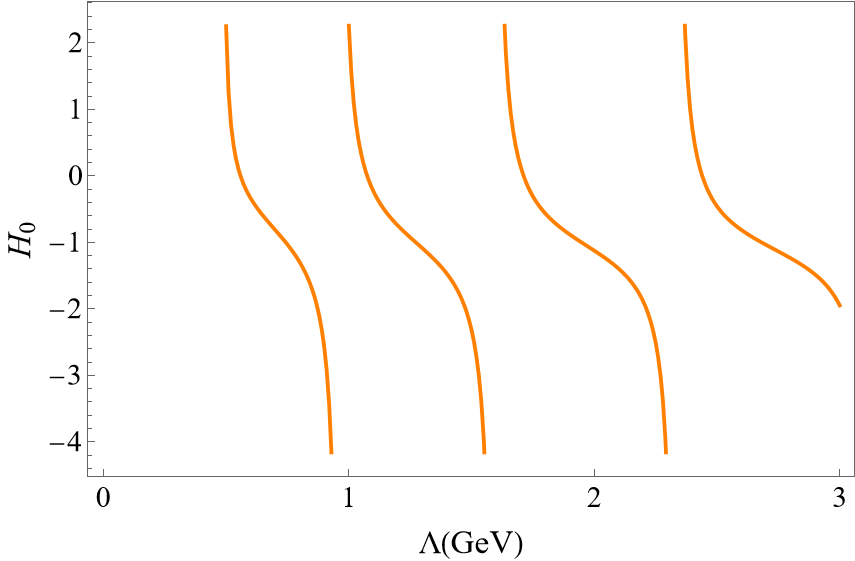}  
  \end{subfigure}
  \begin{subfigure}[b]{0.45\textwidth}
    \includegraphics[width=\textwidth]{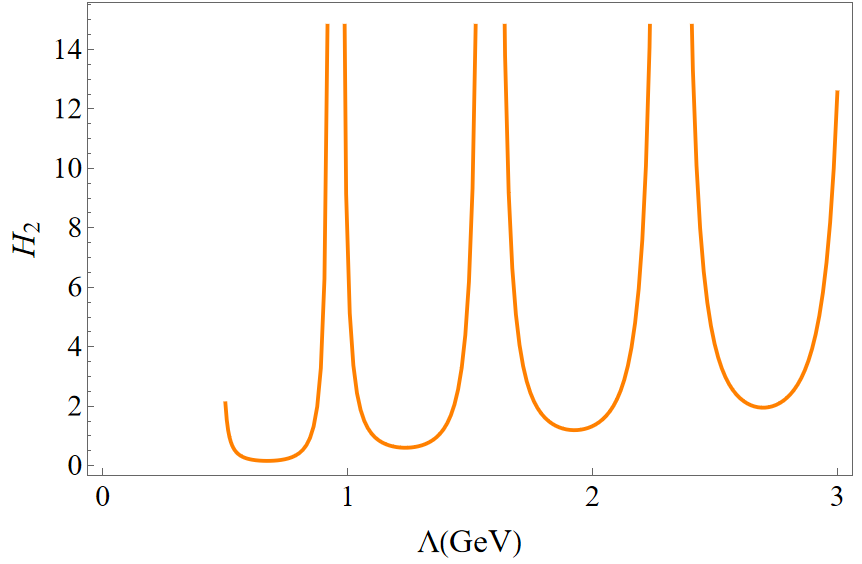}
  \end{subfigure}
  \caption{\justifying Running behavior of three-body couplings. Left
panel: running of $H_{0}$. Right panel: running of $H_{2}$. At $O(p^{2})$,
$H_{0}$ is fit to scattering length of $DD_{s0}^{*}(2317)$ and $H_{2}$
is fit to three-body bound state of $DDK$. }
  \label{fig:3}
\end{figure}

In the infinite volume, we adopt the three-body coupling constants,
i.e., $H_{0}$ and $H_{2}$ to parameterize order by order the short-range
interaction in $DDK$ system. Once the parameters in two-body part
are determined, the core dynamic information of the three-body system
is encoded in $H_{0}$ and $H_{2}$. Therefore, they serve as the
bridge linking the three-body physical observables in the infinite
volume with the lattice spectrum in a finite volume. In order to resolve
the three-body coupling constants, we perform partial wave expansion
on the scattering equation, (\ref{34}). At $S$-wave,
it reads, 
\begin{align}
\mathcal{M}^{(S)}(p,q;E) & =Z^{(S)}(p,q;E)+\int^{\Lambda}\frac{d^{3}k}{(2\pi)^{3}2\omega(k)}Z^{(S)}(p,k;E)\tau(s_{k})\mathcal{M}^{(S)}(k,q;E).
\end{align}
In the infinite volume, Lorentz symmetry is preserved. So in the above
equation, we have chosen rest frame, $P=(E,0,0,0)$ and trivial velocity,
$v^{\mu}=(1,0,0,0)$. Here $p$, $q$, $k$ are the magnitudes of
three-momentum. Dimer invariant mass squared is defined as $s_{k}=(E-\omega_{k})^{2}-k^{2}$.
The projected potential $Z^{(S)}$, e.g., in the channel $(DK)+D\to(DK)+D$,
reads,
\begin{align}
Z_{11}^{(S)}(p,q;E) & =\frac{1}{4pq}\log\frac{\omega_{D}(p)+\omega_{D}(q)+\omega_{K}(p+q)-E}{\omega_{D}(p)+\omega_{D}(q)+\omega_{K}(p-q)-E}\nonumber \\
 & +\frac{H_{0}(\Lambda)}{\Lambda^{2}}+\frac{H_{2}(\Lambda)}{\Lambda^{4}}(s-s_{\text{th}}).
\end{align}
Since there is a physical pole in $DK$ dimer which corresponds to
$D_{s0}^{*}(2317)$, $\sqrt{s_{\text{th}}}$ can be set as the threshold
of $DD_{s0}^{*}(2317)$, that is $2.23954m_{D}$. Below $s_{\text{th}}$,
the pole position of $\mathcal{M}^{(S)}$ is interpreted as three-body
bound state of $DDK$. This bound state is predicted at $\sqrt{s}=2.2266m_{D}$ \cite{Wu:2019cp}. Above $s_{\text{th}}$ and below three-body threshold, $\mathcal{M}^{(S)}$
is proportional to the scattering amplitude of $D$ and $D_{s0}^{*}(2317)$.
Considering wave-function normalization, we can extract the phase
shift of $DD_{s0}^{*}(2317)$. At the threshold, the scattering length
is estimated as $\sim0.45\text{ fm}$ in our calculation \footnote{The two three-body couplings, $H_{0}$ and $H_{2}$ are fitted by
$DD_{s0}^{*}(2317)$'s scattering length and $DDK$'s binding energy.
We can find that the scattering length is $0.4673\text{ fm}$ fm by fitting $DDK$'s
binding energy and setting $H_{2}=0$. So the scattering length is
estimated at the same order, but considered as an independent input
here.}.
Given by $DD_{s0}^{*}(2317)$ scattering length and the $DDK$ bound state,
three-body coupling constants $H_{0}$ and $H_{2}$ can be fit order
by order, e.g., $H_{0}=-0.92453$ and $H_{2}=0.60402$ at $\Lambda=1.25\text{ GeV}$.
We show the running behavior of $H_{0}$ and $H_{2}$ in Fig.\ref{fig:3}. The
logarithmic periodic pattern of Efimov effect can also be observed
in the relativistic kinematics. 

\subsection{Finite volume spectra of \textit{DDK} }
\begin{figure}[p]
    \centering
    \begin{minipage}{0.4\textwidth}
        \centering
        \includegraphics[width=\linewidth, height=0.4\textheight]{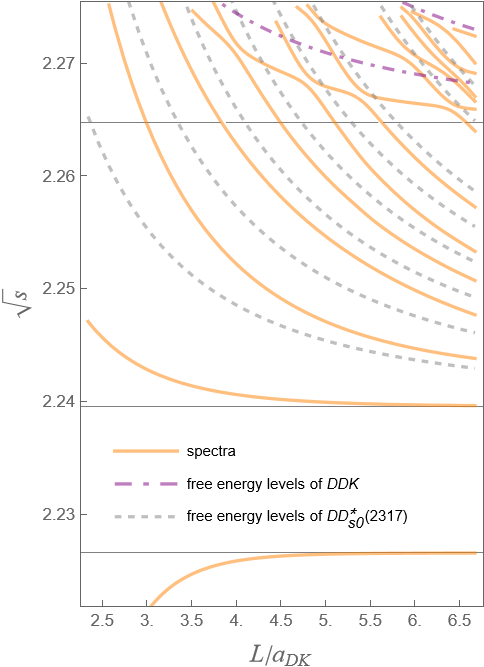}
        \subcaption{a) total momentum: (0,0,0)}
        \label{fig:22a}
    \end{minipage}\hfill
    \begin{minipage}{0.4\textwidth}
        \centering
        \includegraphics[width=\linewidth, height=0.4\textheight]{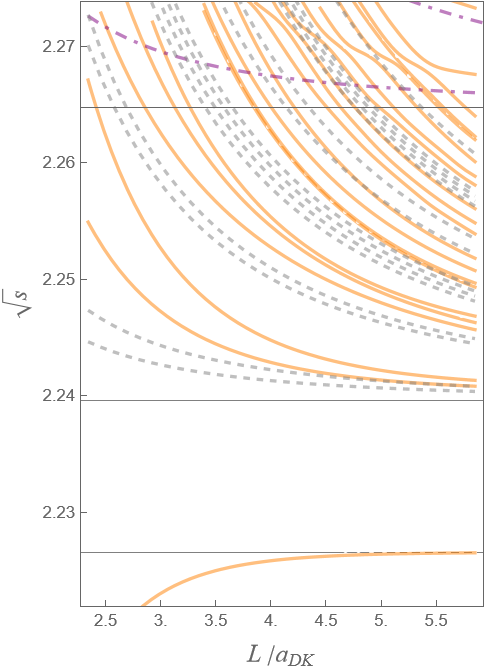}
        \subcaption{b) total momentum: (0,0,1)}
        \label{fig:22b}
    \end{minipage}
    
    \medskip
    
    \begin{minipage}{0.4\textwidth}
        \centering
        \includegraphics[width=\linewidth, height=0.4\textheight]{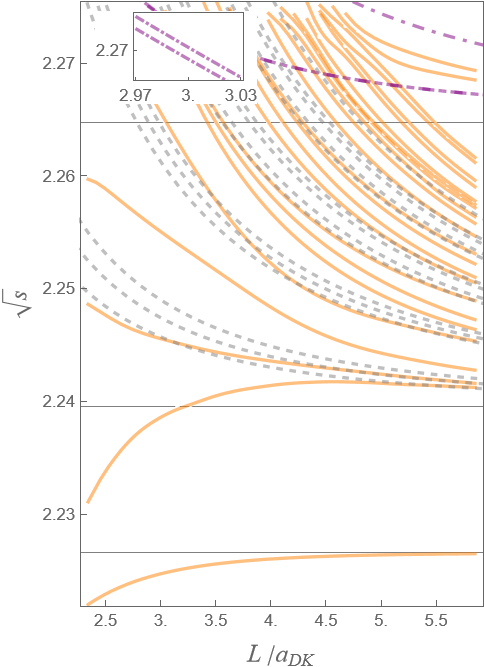}
        \subcaption{c) total momentum: (0,1,1)}
        \label{fig:22c}
    \end{minipage}\hfill
    \begin{minipage}{0.4\textwidth}
        \centering
        \includegraphics[width=\linewidth, height=0.4\textheight]{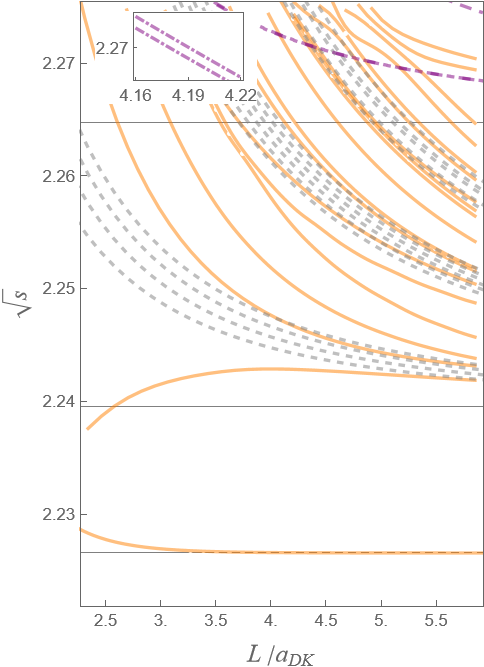}
        \subcaption{d) total momentum: (1,1,1)}
        \label{fig:22d}
    \end{minipage}
    \caption{\justifying Finite volume spectra of $DDK$ three-body system.
a) spectrum in the rest frame; b) the moving frame $(0,0,1)$; c)
$(0,1,1)$; d) $(1,1,1)$. Energy in center of mass $\sqrt{s}$ is
scaled by $m_{D}$. Lattice size $L$ is scaled by $DK$ scattering
length. We present three thresholds in the infinite volume limit denoted
by horizontal black lines. From top to bottom, they are three-body
threshold, $DD_{s0}^{*}(2317)$ two-body threshold and $DDK$ bound
state. All the spectra are present by yellow lines. The purple (gray)
dashed lines are the free energy levels of $DDK$ ($DD_{s0}^{*}(2317)$)
in the four frames. Notice in c) and d), the double-line structures
of $DDK$ free lines are showed in the inset plots.}
    \label{fig:8}
\end{figure}

Based on the three-body coupling constants, we can solve the quantization
condition eq.(\ref{eq:qc}) in both rest frame and moving frames.
We consider the identity representation, that is $A_{1}^{+}$ for
$O_{h}$ in rest frame, i.e., $\mathbf{P}=(0,0,0)$ and
$A_{1}$ for little groups $C_{4v}$, $C_{2v}$, $C_{3v}$ in moving
frames, $(0,0,1)$, $(0,1,1)$, $(1,1,1)$. So the projected quantization
condition is
\begin{align}
\det\Big(\delta_{rr^{'}}-\frac{1}{L^{3}}\frac{\vartheta_{r^{'}}}{|G|}\tilde{\Theta}_{v}(r^{'})Z(r,r^{'};P)\tau((P-k_{r^{'}})^{2})\Big) & =0,
\end{align}
where potential is projected as
\begin{align}
Z(r,r^{'}) & =\sum_{g}Z(gp_{r}^{(0)},q_{r^{'}}^{(0)}).
\end{align}
By solving the quantization condition, we obtain the lattice spectrum
of $DDK$ system for both rest frame and moving frames $(0,0,1)$,
$(0,1,1)$, $(1,1,1)$ in the identity representation (see Fig.\ref{fig:8}).
The energy spectra are given in terms of three-body invariant mass,
$\sqrt{s}=\sqrt{P^{2}}$. 

Although Lorentz invariance is broken in a finite volume, the lattice
spectra show the same limits for large lattice size: 1) three-body
threshold $\sqrt{s_{\text{th}}^{(3)}}=2m_{D}+m_{K}=2.2647m_{D}$; 2)
$DD_{s0}^{*}(2317)$ two-body threshold $\sqrt{s_{\text{th}}^{(2)}}=m_{D}+m_{D_{s0}^{*}(2317)}=2.2395m_{D}$;
3) $DDK$ bound state $\sqrt{s_{\text{B}}}=2.2266m_{D}$ \cite{Wu:2019cp}. They are consistent
with the infinite volume calculations. Below two-body threshold $\sqrt{s_{\text{th}}^{(2)}}$,
we find the predicted $DDK$ bound state in the finite volume spectra
of all the four frames. Between $\sqrt{s_{\text{th}}^{(2)}}$
and three-body threshold $\sqrt{s_{\text{th}}^{(3)}}$, the spectra
can be interpreted as scattering states of $D$ and $D_{s0}^{*}(2317)$
because these spectra can correspond one-to-one with the $DD_{s0}^{*}(2317)$
free lines. Here the $DD_{s0}^{*}(2317)$ free lines is calculated
in the relativistic kinematics, that is 
\begin{align}
\sqrt{s^{(2)}} & =\sqrt{\big(\omega_{D}(\mathbf{p})+\omega_{D_{s0}^{*}(2317)}(\mathbf{P}-\mathbf{p})\big)^{2}-\mathbf{P}^{2}},\\
 & \text{ where }\mathbf{p}\in\left(\frac{2\pi}{L}\right)\mathbb{Z}^{3}.\nonumber 
\end{align}
There is only one kind of two-body states in the spectra, which means
spurious poles has been subtracted successfully, otherwise, we would
see the zigzag lines \cite{fakepole} breaking unitarity. Above $\sqrt{s_{\text{th}}^{(3)}}$,
the three-body scattering states emerge, since they can correspond
one-to-one with $DDK$ free lines, which is calculated by 
\begin{align}
\sqrt{s^{(3)}} & =\sqrt{\big(\omega_{D}(\mathbf{p})+\omega_{D}(\mathbf{q})+\omega_{K}(\mathbf{P}-\mathbf{p}-\mathbf{q})\big)^{2}-\mathbf{P}^{2}},\\
 & \text{ where }\mathbf{p},\mathbf{q}\in\left(\frac{2\pi}{L}\right)\mathbb{Z}^{3}.\nonumber 
\end{align}
To see this more clearly, we can find that for moving frames $(0,1,1)$
and $(1,1,1)$, there are two lowest $DDK$ free lines very close
to each other (Fig.\ref{fig:22c} - \ref{fig:22d}), consequently, the three-body scattering
states are also present in double lines. In the region above $\sqrt{s_{\text{th}}^{(3)}}$,
both the $DD_{s0}^{*}(2317)$ and $DDK$ free lines coexist for all
frames, thus avoided level crossing can be observed in the finite
volume spectra. It reflects the superposition of two-body $DD_{s0}^{*}(2317)$
and three-body $DDK$ pictures, which is similar to two-body scenario
with coupled channels. 

\begin{figure}[h]
\centering
\includegraphics[height=0.4\textheight]{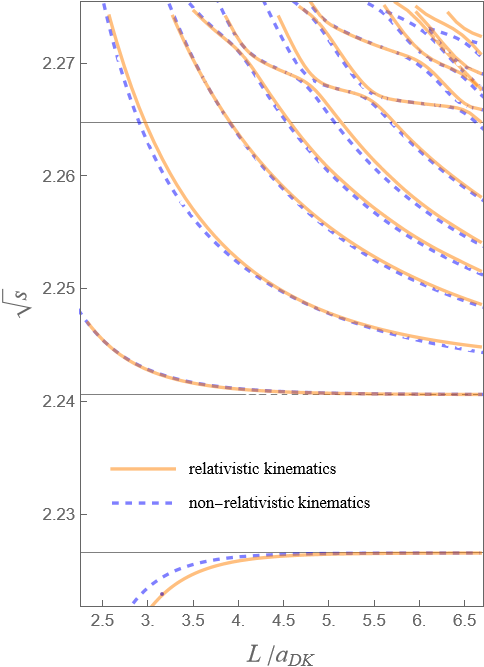}
\caption{\justifying Comparison of spectra with non-relativistic and relativistic kinematics. Energy in center of mass $\sqrt{s}$ is scaled by $m_{D}$. Lattice size $L$ is scaled by $DK$ scattering length. The spectra with non-relativistic kinematics is denoted by blue dashed lines  \cite{ddk}. In order to do the comparison, the parameters used for Lorentz invariant spectra (denoted by yellow lines) are adjust to match \cite{ddk}. The comparison is done in the rest frame, where the two spectra are consistent very well.}
\label{fig:23}
\end{figure}

\begin{figure}[p]
    \centering
    \begin{minipage}{0.4\textwidth}
        \centering
        \includegraphics[width=\linewidth, height=0.4\textheight]{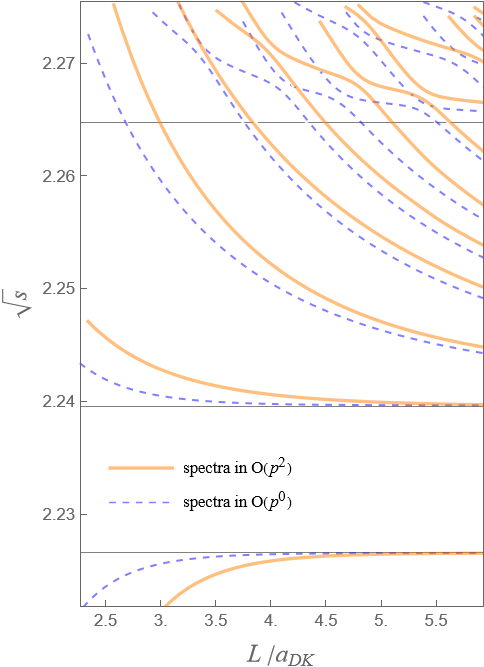}
        \caption*{\centering a)  total momentum: (0,0,0)}
    \end{minipage}\hfill
    \begin{minipage}{0.4\textwidth}
        \centering
        \includegraphics[width=\linewidth, height=0.4\textheight]{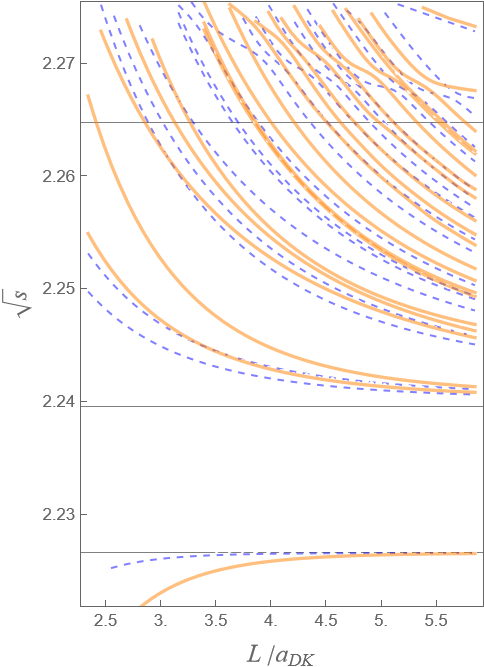}
        \caption*{\centering b)  total momentum: (0,0,1)}
    \end{minipage}
    
    \medskip
    
    \begin{minipage}{0.4\textwidth}
        \centering
        \includegraphics[width=\linewidth, height=0.4\textheight]{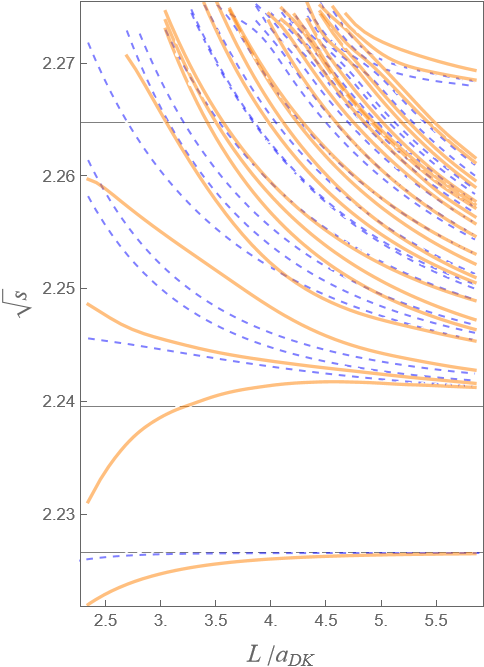}
        \caption*{\centering c)  total momentum: (0,1,1)}
    \end{minipage}\hfill
    \begin{minipage}{0.4\textwidth}
        \centering
        \includegraphics[width=\linewidth, height=0.4\textheight]{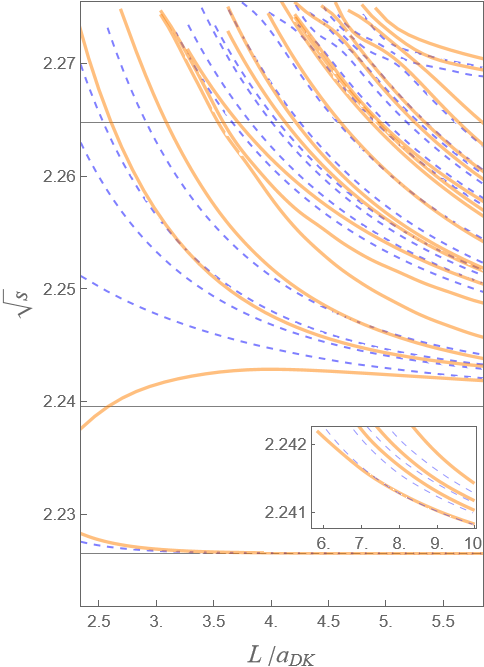}
        \caption*{\centering d)  total momentum: (1,1,1)}
    \end{minipage}
    
    \caption{\justifying Comparison of spectra in $O(p^{0})$ and $O(p^{2})$.
Energy in center of mass $\sqrt{s}$ is scaled by $m_{D}$. Lattice
size $L$ is scaled by $DK$ scattering length. The spectra in $O(p^{0})$
is denoted by blue dashed lines, and $O(p^{2})$ by yellow lines.
They have the same qualitative structure and exhibit convergence in
the limit of large lattice size.}
    \label{fig:9}
\end{figure}

We plot a comparison between the relativistic and non-relativistic \cite{ddk}
cases in the rest frames (Fig.\ref{fig:23}). It shows that for the near-threshold
region of the $DDK$ system, the non-relativistic approximation is
applicable. Of course, in the moving frame, the increase in total
momentum makes the effects of relativistic kinematics more significant.
Therefore, the current work has broader applicability. We also plot
the comparison between spectra at $O(p^{0})$ and $O(p^{2})$ in all
four frames (Fig.\ref{fig:9}). They have the same qualitative structure and
exhibit convergence in the limit of large lattice size, which demonstrates
that the power counting of our formalism is correct.


\section{ Conclusion}\label{sec4}
We have successfully produced the lattice spectra of $DDK$ three-body
system within a Lorentz covariant formalism. It allows us to generate
spectra (Fig.\ref{fig:8}) in various moving frames, i.e., $(0,0,0)$,
$(0,0,1)$, $(0,1,1)$ and $(1,1,1)$. The infinite volume limits,
$DDK$ three-body threshold, $DD_{s0}^{*}(2317)$ two-body threshold
and $DDK$ bound state \cite{Wu:2019cp}, are present properly in all four
frames. The scattering states correspond one-to-one with two kinds
of free states, i.e., $DDK$ three-body free lines and $DD_{s0}^{*}(2317)$
two-body free lines within Lorentz covariant formalism. Avoided level
crossing is also observed above three-body threshold. In the rest
frame, our calculation is consistent with the non-relativistic spectra
\cite{ddk} (Fig.\ref{fig:23}), while in the moving frames,
as the increasing of total momenta, relativistics kinematics becomes
more significant, making the Lorentz covariant extension in this paper
more valuable. The calculations are carried at both $O(p^{0})$ and
$O(p^{2})$ in all four frames (Fig.\ref{fig:9}). The power
counting in NREFT turns to be correct since the spectra of $O(p^{0})$
and $O(p^{2})$ have the same qualitative structure and exhibit convergence
in the limit of large lattice size. 

In order to build up the lattice spectra in a Lorentz covariant manner,
the particle-dimer formalism of $DDK$ system with relativistic kinematics
is established. Dimer propagators are analyticly continuated below
threshold for the three-body calculation. Up to $O(p^{2})$, the spurious
pole \cite{fakepole} in both $DK$ and $DD$ dimers are subtracted
in a Lorentz-invariant manner. We demonstrate the analyticity and
unitarity can be preserved order by order (Fig.\ref{fig:xy}).
In three-body sector, we introduce $H_{0}$ and $H_{2}$ at $O(p^{0})$
and $O(p^{2})$ to parameterized the three-body interaction of $DDK$.
By fitting to $DD_{s0}^{*}(2317)$ scattering length and $DDK$ three-body
bound state predicted by \cite{Wu:2019cp}, the logarithmic periodical running
behavior of $H_{0}$ and $H_{2}$ are present properly (Fig.\ref{fig:3}).
The Lorentz covariant quantization condition equipped with these three-body
couplings generates lattice spectra of $DDK$ in various moving frame.
Since the three-body physical quantities, i.e., $DD_{s0}^{*}(2317)$
scattering length and $DDK$ three-body bound state are both theoretical
assumptions, our lattice spectra is a prediction to test these premises.
In another word, when the $DDK$ spectra is obtained on lattice, our
formalism can be applied to extract three-body couplings order by
order via quantization condition. These low-energy contants derived
from the QCD first principles can predict reliably the three-body
observables in $DDK$ system via Lorentz-invariant particle-dimer
formalism in the infinite volume. Additionally, we can improve the
current work by introducing more detailed two-body parameters, e.g.,
higher shape parameters in $DK$ and $DD$ scattering channels, as
well as three-body coupling beyond $O(p^{2})$. Since the relativsitic
kinematics are involved, power counting is applicable for all calculations
under the threshold of relativistic dynamics. These works are important
as long as the sufficient lattice data in the future simulation is
supplied, which is expected.  On the
other hand, applying our work on $DDK$ to $DD\pi$ or $D\bar{D}\pi$
system can serve as an alternative lattice study for $T_{cc}(3875)$ \cite{Hansen2024}
or $X(3872)$. 
\section*{Acknowledge}
The authors would like to thank Akaki Rusetsky for interesting discussions. The work of Q.-C.-X., J.-Y.P. and J.-J.W. was supported by National
Natural Science Foundation of China (NSFC) under Grants No. 12130511,
12175239, and by the National Key R\&D Program of China under Contract
No. 2020YFA0406400.

\end{document}